%% file: main.tex
\documentclass{aa}  

\usepackage{graphicx}
\usepackage{txfonts}
\usepackage[hidelinks]{hyperref}
\usepackage{physics}
\usepackage[utf8]{inputenc}
\usepackage[english]{babel}
\usepackage[T1]{fontenc}
\usepackage{lmodern}
\usepackage{microtype}
\usepackage{booktabs}
\usepackage{bm}
\usepackage{graphicx}
\usepackage{amsfonts}
\usepackage{amsmath}
\usepackage{amssymb}
\usepackage{listings}
\usepackage{subfigure}
\usepackage{fancyhdr}
\usepackage{mathtools}
\usepackage{natbib}
\usepackage{comment}
\usepackage{orcidlink}
\usepackage{longtable}

\newcommand{\kms}{\mathrm{km/s}}
\newcommand{\single}{\mathcal{S}}
\newcommand{\binary}{\mathcal{B}}
\newcommand{\psingle}{p_\mathrm{single}}
\def\srv{$\sigma_\mathrm{1D}$}
\def\fbin{$f_\mathrm{bin}$}
\def\pcutoff{$P_\mathrm{cutoff}$}

\usepackage{color}

\bibpunct{(}{)}{;}{a}{}{,} 

\begin{document}

\titlerunning{Non-parametric binary identification with single-epoch spectroscopy}
\authorrunning{S.~Rinaldi \& M.~C.~Ramírez-Tannus}
\title{Non-parametric identification of single-lined binary candidates in young clusters using single-epoch spectroscopy}

\author{
Stefano~{Rinaldi}\inst{1,2}\thanks{E-mail: stefano.rinaldi@uni-heidelberg.de}\,\orcidlink{0000-0001-5799-4155}
\and
María~Claudia~{Ramírez-Tannus}\inst{3}\thanks{E-mail: ramirez@mpia.de}\,\orcidlink{0000-0001-9698-4080}
}

\institute{
Institut für Theoretische Astrophysik, ZAH, Universität Heidelberg, Albert-Ueberle-Str. 2, 69120 Heidelberg, Germany
\and
Dipartimento di Fisica e Astronomia ``G. Galilei'', Università di Padova, Via Marzolo 8, 35121 Padova, Italy
\and
Max-Planck Institut für Astronomie, Königstuhl 17, 69117 Heidelberg, Germany
}

\date{Received \today; accepted XXX}
 
\abstract
{}
{Binarity plays a crucial role in star formation and evolution. Consequently, identifying binary stars is essential to deepen our understanding of these processes. We propose a method to investigate the observed radial velocity distribution of massive stars in young clusters with the goal of identifying binary systems.}
{We reconstruct the radial velocity distribution using a three-layers hierarchical Bayesian non-parametric approach: non-parametric methods are data-driven models able to infer arbitrary probability densities under minimal mathematical assumptions. When applying our statistical framework, it is possible to identify variable stars and binary systems because these deviate significantly from the expected intrinsic Gaussian distribution for radial velocities.}
{We test our method with the massive star forming region within the giant H$_\mathrm{II}$ region M17. We are able to confidently identify binaries and variable stars with as little as single-epoch observations. The distinction between variable and binary stars improves significantly when introducing additional epochs.}
{}


\maketitle
\section{Introduction}

In the last decades, it has become clear that multiplicity is a fundamental aspect of stellar formation and evolution, especially towards higher masses \citep[see][~and references therein]{Offner:2023}.  Massive stars are preferentially observed in binary systems, close enough to interact with their companions during their lifetime. The latter is shown by observations of Milky Way clusters \citep[][]{2012Sci...337..444S, sana:2013, 2015A&A...580A..93D, 2015ApJ...810...61M, 2017IAUS..329...89B, 2017ApJS..230...15M, 2022A&A...658A..69B} as well as in the LMC \citep[][]{2017A&A...598A..84A, 2021MNRAS.507.5348V}. The presence of close companions affects all stages of the stellar life, from the pre-main sequence phase \citep[][]{2024NatAs...8..472L} to the end-of-life explosions, potentially introducing new physics during the stellar life \citep[e.g.][]{2019Natur.574..211S, 2024Sci...384..214F} and affecting the properties and orbital evolution of double-compact objects, which may eventually become the progenitors of gravitational wave sources \citep[][]{1991ASIC..342..125B, 2007ARA&A..45..481Z,  2012ARA&A..50..107L, 2013ApJ...764..166D, 2014prpl.conf..149T}.

Due to the proximity of the stars in close binary systems, these systems are usually not possible to be resolved spatially. Therefore, the best way of identifying them is in velocity space via spectroscopic measurements. Close binaries are detected through periodic Doppler shifts of the photospheric lines in their spectra \citep[e.g.,][]{sana:2013, 2015A&A...580A..93D, 2014ApJS..213...34K}. 
Spectroscopic binaries detected through multi epoch spectroscopy are usually divided in two categories, SB1 and SB2. SB2 refers to double-lined spectroscopic binaries, in which both components are visible in the spectra, whereas SB1 refers to single-line spectroscopic binaries with only one component of the system is visible in the spectrum. The binary nature of the system is determined via radial velocity shifts. 
Several surveys aimed at identifying massive binary stars have been carried out both in the Milky Way and in lower metallicity environments such as the LMC and SMC \citep[e.g.,][]{sana:2013, 2015A&A...580A..93D, 2014ApJS..213...34K, 2017A&A...598A..84A, 2021MNRAS.507.5348V}. The minimum number of epochs covered by such surveys is typically three, with the second epoch a few days after the first one and the third a few months after. This allows to efficiently detect binaries with periods up to 10 days as well as those with periods up to 6 months. However, three epochs are usually not sufficient if one wants to get accurate orbital solutions: some surveys extent to tens of epochs per object \citep[e.g.][]{sana:2013}. 

An indirect method to estimate the amount of close binaries in a cluster was proposed by \citet{2017A&A...599L...9S} and \citet{2021A&A...645L..10R}. In these studies, the authors measured the radial velocity dispersion (\srv) of single epoch observations, which in low density clusters is strongly dominated by the orbital properties of the binary population. The multiplicity properties of a cluster can be quantified by comparing the measured \srv\ value with that resulting from Monte Carlo population synthesis, where each parent population is characterised by different underlying multiplicity properties. This method has the advantage of allowing to get an idea of the underlying multiplicity properties of a cluster with a single epoch observation, reducing considerably the observing time with respect to a multi epoch approach. However, there are several underlying multiplicity properties that have an effect on the \srv\ of a cluster and those cannot be individually determined based on the \srv\ only. For example, \citet{2017A&A...599L...9S} and \citet{2021A&A...645L..10R} show that the low \srv\ measured for the young cluster M17 can either be explained either by a low \fbin\ or by a large \pcutoff\ with respect to other, slightly older Galactic clusters. In order to determine which of the two scenarios causes the low \srv\ observed it is necessary to determine the actual \fbin\ of the cluster. 

In this paper we propose a statistical framework based on a Bayesian non-parametric method to efficiently identify binary candidates through single epoch spectroscopy. 
Non-parametric methods are powerful tools to infer probability distributions without being committal to any specific functional form and, in a certain sense, \emph{let the data speak for themselves}. The name `non-parametric methods' might be misleading: in fact, these models have an infinite number of parameters that can be used to accommodate arbitrary probability densities without the need for any fine-tuning. Such flexibility comes with the cost that it is not possible to give a direct physical interpretation of the inferred parameters, thus any interpretation of the reconstructed features of the distribution must be done a posteriori. The framework presented in this work, being applicable to single-epoch observations, can reduce considerably the observing time required in comparison to the multi epoch approach, the flexibility of non-parametric methods will allow us to account both for outliers (such as stars in binary systems with limited epochs) and deviations from the expected Gaussian shape of the underlying distribution.
Most importantly, this flexible characterisation of the radial velocity distribution allows us to identify the systems that are worth following up with multi-epoch spectroscopy, opening for a further optimisation of the available observing time usage. 

The paper is structured as follows: in Section~\ref{sec:methods} we present the statistical framework that allows us to detect binary stars based on single epoch spectroscopy. Section~\ref{sec:simulations} presents the application of this framework to a set of simulated observations to assess its applicability, whereas in Section~\ref{sec:test_case} we apply it to a set of observations within the massive star forming region M17 and compare our binary identification to that of \citet{ramirez:2024}, who identified binary stars based on multi-epoch spectroscopy. In Section~\ref{sec:conclusion} we discuss and conclude this work.  

\section{Methods}\label{sec:methods}
In this section, we will present the statistical framework used to infer the radial velocity distribution and to identify the spectroscopic binary systems in the cluster. 

The main goal of this paper is to identify spectroscopic variable objects rather than reconstructing the intrinsic distribution of radial velocities: in what follows we will always implicitly refer to observed distributions, thus not including a treatment of selection effects in this framework. The only assumption made in this sense is that the detection probability of either a star or a binary system is independent of its centre-of-mass (CM) radial velocity.

\subsection{Radial velocity distribution}
Determining whether an observed object in a cluster is a single star or a binary system having measured its radial velocity is by all means a classification problem, addressed in this paper in a Bayesian framework by stating the two following hypotheses, formulated as propositions:
\begin{itemize}
\item[$\single$:] The object is a single star;
\item[$\binary$:] The object is variable or part of a binary system.
\end{itemize}
In principle, the radial velocity of an observed star can fluctuate also because of the intrinsic variability of the star itself. Stellar variability can cause radial velocity variations of up to 20~$\kms$ \citep[][]{2009A&A...507.1585R}.
From a statistical point of view, however, the two phenomena (intrinsic variable and binary component) are indistinguishable. Therefore, in what follows we will use `binary candidate' as an umbrella term to denote all the possible phenomena that may cause variability in radial velocity measurements. Moreover, we will make use of the word `population' in a statistical sense, meaning that the objects have some properties that are drawn from the same distribution -- in the case of this paper, the radial velocity.
This is different to the common use of the same word in the astrophysical context, when one refers to a population as a group of objects sharing the same age, proper motion and position in space.

For thermalised systems such as clusters, dynamical interactions are expected to produce a radial velocity distribution for the CM of the systems\footnote{Here we consider single stars as systems composed by one object only, with their CM coinciding with the one of the star.} which is Gaussian in shape and with a standard deviation (often called \emph{velocity dispersion}) $\sigma_V$:
\begin{equation}\label{eq:vcm-thermalised}
    p(v_{cm}) = \mathcal{N}(v_{cm}|V_0,\sigma_V)\,.
\end{equation}
Single stars, since they coincide with their own CM, follow the same distribution:
\begin{equation}\label{eq:rv-single}
p(v_r|\single) = \int \delta(v_r-v_{cm})p(v_{cm})\dd v_{cm} = \mathcal{N}(v_r|V_0,\sigma_V)\,.
\end{equation}

For stars in binary systems, however, we are observing a combination of both the movement of the star around the CM of the binary and the motion of the CM itself:
\begin{multline}\label{eq:rv-binary}
    p(v_r|\binary) = \int p(v_r|v_{cm})p(v_{cm}) \dd v_{cm} = \\ = \int p(v_r|v_{cm})\mathcal{N}(v_{cm}|V_0,\sigma_V)\dd v_{cm}\,,
\end{multline}
The distribution $p(v_r|v_{cm})$ stems from the fact that the orbital parameters -- in particular the period and phase -- of the system at the moment of the observation are unknown\footnote{The same thing happens also for intrinsic variables, whereas in this second case the distribution $p(v_r|v_{cm})$ is caused by the chaoticness of the underlying phenomenon.}.

Moreover, individual objects are often tracked through different epochs: whereas for single stars multiple observations are scattered around the intrinsic radial velocity of the star, for a binary component multi-epoch observations tracks different phases of the orbit. Therefore, for multi-epoch observations of a binary component, observing the system at random times is equivalent to drawing samples from $p(v_r|v_{cm})$\footnote{Strictly speaking, this holds if the (random) interval $\Delta T$ between subsequent observations at least comparable with the binary period $P$: $\Delta T \gtrsim P$.}.

Due to the presence of uncertainties and errors, the measured radial velocity $\hat v_r$ is scattered around the true value $v_r$ with a typical dispersion corresponding to the measurement uncertainty $\delta_r$. Making use of the Maximum Entropy Principle \citep{jaynes:2003}, we assume a Gaussian distribution for $p(v_r|\hat v,\delta)$:
\begin{equation}\label{eq:measurement-likelihood}
    p(v_r|\hat v,\delta) = \mathcal{N}(v_r| \hat v,\delta)\,.
\end{equation}

In what follows, we will assume that our observed catalogue contains $N$ objects, that for each of the $N$ objects we have $n_i$ epochs and for each epoch we have a measured radial velocity $\hat v_{j}$ and a measurement uncertainty $\delta_{j}$. We will denote with $O_i = \{(\hat v_{1},\delta_{1}),\ldots,(\hat v_{n_i},\delta_{n_i})\}$ the ensemble of measurements for the $i-$th star and with $\mathbf{O} = \{O_1,\ldots,O_N\}$ the catalogue of stars. 

In principle, one could model the CM radial velocity distribution as a simple Gaussian distribution and then infer its parameters -- which is the implicit assumption behind the use of the cluster's velocity dispersion as proxy for the presence of binary system -- using a three-layers hierarchical framework, $p(v_{cm}) - p(v_r) - p(\hat v)$. Using the velocity dispersion obtained from the observations to determine whether or not the catalogue contains binaries, albeit not wrong \emph{per se}, doesn't fully capture all the features of the CM radial velocity distribution.
In particular, in all these situations in which the number of epochs per object is small (one or two), the limited number of observations for the radial velocity of the binary system does not allow to properly characterise and account for the orbit of the star around the CM (Eq.~\ref{eq:rv-binary}), thus biasing the inferred velocity dispersion.

Properly characterising all the features present in $p(v_{cm})$ is crucial not only to assess the presence of binaries, but most importantly to guide the development of models aiming at reproducing this distribution. The basic idea we put forward in this work is to approximate the distribution of the CM radial velocity $p(v_\mathrm{cm})$ of the considered cluster with an infinite mixture of Gaussian distributions:
\begin{equation}\label{eq:dpgmm}
    p(v_{cm}) \simeq \sum_k^\infty w_k \mathcal{N}(v_{cm}|\mu_k, \sigma_k)\,.
\end{equation}
This model, called Dirichlet process Gaussian mixture model (DPGMM) \citep{escobar:1995} when equipped with a Dirichlet process prior to handle the countably infinite number of Gaussian components, is a reliable approximant for continuous probability densities with support on $\mathbb{R}$ \citep{nguyen:2020}. The countably infinite parameters for such distribution, denoted with $\Theta$, are the weights $\mathbf{w} = \{w_1,w_2,\ldots\}$, means $\boldsymbol\mu = \{\mu_1,\mu_2,\ldots\}$ and standard deviations $\boldsymbol\sigma = \{\sigma_1,\sigma_2,\ldots\}$: $\Theta =\{\mathbf{w}, \boldsymbol\mu, \boldsymbol\sigma\}$. These can be inferred using the available observations $\mathbf{O}$:
\begin{multline}
    p(\Theta|\mathbf{O}) = \int p(\Theta|\mathbf{v}_{cm})p(\mathbf{v}_{cm}|\mathbf{O}) \dd \mathbf{v}_{cm} \\ = \int p(\Theta|\mathbf{v}_{cm})\prod_i^N p(v_{cm,i}|O_i) \dd \mathbf{v}_{cm}\,.
\end{multline}
Here we denoted with $\mathbf{v}_{cm} = \{v_{cm,1},\ldots,v_{cm,N}\}$ the set of (unknown) CM radial velocities and made use of the fact that individual objects are independent realisations from the same distribution (independent and identically distributed - i.i.d.). In general, writing a functional form for $p(v_{cm,i})$ requires the knowledge of whether the object is a single star or the component of a binary system. Moreover, in the latter case, one should also model $p(v_{cm,i})$ making use of its orbital parameters. Here we address the issue modelling each $p(v_{cm,i})$, once again, as a DPGMM. The parameters of each of these mixtures will be denoted with $\theta_i = \{\mathbf{w}_i, \boldsymbol\mu_i, \boldsymbol\sigma_i\}$.
Under this assumption, we get
\begin{equation}\label{eq:outer-dpgmm}
    p(\Theta|\mathbf{O}) = \int p(\Theta|\mathbf{v}_{cm})\prod_i^N p(v_{cm,i}|\theta_i) p(\theta_i|O_i)\dd \theta_i \dd \mathbf{v}_{cm}\,.
\end{equation}
This approach solves both issues at once: in the case of a single star, the DPGMM approximates the delta function in Eq.~\eqref{eq:rv-single} with a very narrow distribution, whereas for binary systems we get an approximant for $p(v_{cm,i})$ using multiple epochs (if available) as samples from this distribution\footnote{If only very few epochs are available, the reconstructed distribution will not be a reliable approximant for $p(v_{cm,i})$. This is not an issue for the method presented in this paper, since the non-parametric method used to model $p(v_{cm})$ will account for such fluctuations.}. This two-layer hierarchical model is the same presented in \citet{rinaldi:2022:hdpgmm}, where we introduce `a Hierarchy of Dirichlet process Gaussian mixture models', or (H)DPGMM for short. This method, originally developed for the inference of the black hole mass distribution \citep{rinaldi:2022:hdpgmm,rinaldi:2024:m1qz}, has proven itself extremely flexible and suitable for a variety of different problems \citep{rinaldi:2022:skyloc,rallapalli:2023,rinaldi:2024:evidence}. 

Making use of Eq.~\eqref{eq:measurement-likelihood} and denoting with $\mathbf{v}_{r,i} = \{v_{r,1}, \ldots, v_{r,n_i}\}$, the posterior distribution for the parameters of each individual DPGMM reads:
\begin{multline}\label{eq:inner-dpgmm}
    p(\theta_i|O_i) = \int p(\theta_i|\mathbf{v}_{r,i})p(\mathbf{v}_{r,i}|O_i) \dd\mathbf{v}_{r,i}\\ = \int p(\theta_i|\mathbf{v}_{r,i})\prod_j^{n_i}\mathcal{N}(v_{r,j}|\hat v_{i,j},\delta_{i,j}) \dd\mathbf{v}_{r,i}\,.
\end{multline}
The individual Gaussian distribution used to model the individual measurement uncertainty can be interpreted as a single-component Gaussian mixture with known parameters: with this in mind, the resemblance between Eq.~\eqref{eq:outer-dpgmm} and Eq.~\eqref{eq:inner-dpgmm} is evident. In practice, this model can be explored first reconstructing the CM radial velocity probability distribution for the individual objects using Eq.~\eqref{eq:inner-dpgmm} and then combining these reconstructions together to get an approximant for $p(v_{cm})$ using Eq.~\eqref{eq:outer-dpgmm}.
The framework presented here can be thought, in fact, as a hierarchy of (H)DPGMMs.

\subsection{Binary candidate identification}\label{sec:binary_identification}
We now turn our attention to the problem of assessing whether an observed star is single or a spectroscopic binary candidate. Following \citet{rinaldi:2022:hdpgmm} and \citet{rinaldi:2022:skyloc}, we will take an approach modelled on the Gibbs sampling scheme \citep{neal:2000,gorur:2010}, thus based on conditional distributions. The Gibbs approach explores the parameter space iteratively fixing the values of a subset of the parameters (in the simplest case, all but one) and updating the others making use of the conditional probability distribution.
In a nutshell, assuming a partition of the parameter space $\theta = \{\alpha, \beta\}$, a pseudo-algorithm for the Gibbs sampling scheme looks like the one sketched in Listing~\ref{gibbs-code}.
\begin{lstlisting}[label=gibbs-code, caption = Gibbs sampling pseudo-algorithm, frame=tb, mathescape = true, escapeinside={(*}{*)}]
initialise $\alpha$, $\beta$
iterate:
    iterate K times:
        draw $\alpha$ from $p(\alpha|\beta)$
        update $\alpha$
        draw $\beta$ from $p(\beta|\alpha)$
        update $\beta$
    save $(\alpha,\beta)$
\end{lstlisting}
This approach is useful in all these situations in which the joint probability distribution $p(\alpha,\beta)$ is either too expensive or even impossible to compute.

For the purpose of this derivation, we introduce a set of $N$ auxiliary variables $\mathbf{z} = \{z_1,\ldots,z_N\}$: these variables label each object either as a single star ($z_i = \single$) or part of a binary system ($z_i = \binary$). In what follows, we will denote with $\mathbf{z}_{-i}$ the set of variables $\mathbf{z}$ without the $i$-th component. Moreover, we will make use of the following two assumptions:
\begin{itemize}
    \item The CM radial velocity distribution is Gaussian with standard deviation $\sigma_V$;
    \item The Gaussian feature is the most prominent feature of the CM radial velocity distribution.
\end{itemize}
These two assumptions allow us to characterise $p(v_{cm})$ as in Eq.~\eqref{eq:vcm-thermalised}, fixing the mean $V_0$ in correspondence of the maximum of the inferred DPGMM approximant for this distribution. $\sigma_V$ is then estimated minimising the Jensen-Shannon distance \citep{lin:1991} $\mathrm{JS}(p,q)$ between the median reconstructed $p(v_{cm})$ and a normal distribution with mean $V_0$ and variance $\sigma$:
\begin{equation}\label{eq:jsdistance}
    \sigma_V = \min_{\sigma}\Bigg( \mathrm{JS}\Big(p(v_{cm}), \mathcal{N}(v_{cm}|V_0, \sigma)\Big)\Bigg)\,.
\end{equation}

Assuming that we already know all the values of $\mathbf{z}_{-i}$, the probability distribution for $z_i$ reads
\begin{multline}\label{eq:singlebinaryprob}
    p(z_i = \single/\binary|O_i,\mathbf{z}_{-i}) \\ = \frac{p(O_i|z_i = \single/\binary)p(z_i = \single/\binary|\mathbf{z}_{-i})}{p(z_i = \single|O_i,\mathbf{z}_{-i}) + p(z_i = \binary|O_i,\mathbf{z}_{-i})}\,,
\end{multline}
assuming that every observation $O_i$ is independent of the others. 
The second term of the numerator is the probability of the observed object to be a single star or a binary component solely based on on the number of objects already assigned to each of the two categories. Denoting with $N_\single$ the number of objects with $z_i = \single$, with  $N_\binary$ its equivalent with $z_i = \binary$ and with $w_\single$ the single star fraction, we can write
\begin{multline}\label{eq:multinomial-explicit}
    p(\mathbf{z}) = \int p(\mathbf{z}|w_\single)p(w_\single)\dd w_\single \\ = \int w_\single^{N_\single}(1-w_\single)^{N_\binary}p(w_\single)\dd w_\single\,. 
\end{multline}
Making use of a symmetric Beta distribution with shape parameter $\beta$ as prior on $w_\single$,
\begin{equation}\label{eq:betadist}
    p(w_\single|\beta) = \frac{\Gamma(\beta)}{\qty(\Gamma(\beta/2))^2}w_\single^{\beta/2-1}(1-w_\single)^{\beta/2-1}\,,
\end{equation}
where $\Gamma(\beta)$ is the Gamma function, the integral in Eq.~\eqref{eq:multinomial-explicit} can be carried out analytically:
\begin{equation}\label{eq:multinomial-implicit}
    p(\mathbf{z}) = \frac{\Gamma(\beta)\Gamma(N_\single + \beta/2)\Gamma(N_\binary + \beta/2)}{\Gamma(N+\beta)\qty(\Gamma(\beta/2))^2}\,.
\end{equation}
The probability we are trying to compute can be rewritten as
\begin{equation}
    p(z_i = \single/\binary|\mathbf{z}_{-i}) = \frac{p(z_i = \single/\binary, \mathbf{z}_{-i})}{p(\mathbf{z}_{-i})}\,.
\end{equation}
Making use of Eq.~\eqref{eq:multinomial-implicit}, this probability takes a simple form: 
\begin{equation}\label{eq:p_classes_single_binary}
    p(z_i = \single/\binary|\mathbf{z}_{-i}) = \begin{cases}
        \frac{N_\single + \beta/2}{N+\beta} \qif z_i = \single\\
        \frac{N_\binary + \beta/2}{N+\beta} \qif z_i = \binary
    \end{cases}\,.
\end{equation}

We now turn our attention to the first term of the numerator in Eq.~\eqref{eq:singlebinaryprob}. This quantity, under the assumption $\binary$, reads
\begin{multline}\label{eq:p_binary}
    p(O_i|z_i = \binary) = \prod_j^{n_i}\int p(\hat v_{i,j},\delta_{i,j}|v_{r,j})p(v_{r,j}|\binary) \dd v_{r,j} \\ \propto \prod_j^{n_i}\int \mathcal{N}(v_{r,j}|\hat v_{i,j},\delta_{i,j})p(v_{r,j}|z_i = \binary) \dd v_{r,j}\,.
\end{multline}
The term $p(v_{r,j}|z_i = \binary)$ is given in Eq.~\eqref{eq:rv-binary} and can be approximated with the DPGMM reconstruction obtained from Eq.~\eqref{eq:inner-dpgmm}.
On the other hand, for a single star, the same quantity becomes
\begin{multline}\label{eq:p_single}
    p(O_i|z_i = \single) = \int\prod_j^{n_i} p(\hat v_{i,j},\delta_{i,j}|v_{r,j})p(v_{r,j}|\single) \dd v_{r,j} \\= \int \prod_j^{n_i} p(\hat v_{i,j},\delta_{i,j}|v_{r,j})p(v_{r,j}|v_{cm})p(v_{cm}) \dd v_{r,j} \dd v_{cm}
    \\ \propto \int \prod_j^{n_i} \mathcal{N}(v_{r,j}|\hat v_{i,j},\delta_{i,j})\delta(v_{r,j}-v_{cm})\mathcal{N}(v_{cm}|V_0,\sigma_V) \dd v_{r,j} \dd v_{cm}
    \\ \propto \int \prod_j^{n_i} \mathcal{N}(v_{cm}|\hat v_{i,j},\delta_{i,j})\mathcal{N}(v_{cm}|V_0,\sigma_V) \dd v_{cm}\,,
\end{multline}
where we made use of Eq.~\eqref{eq:rv-single} to specify $p(v_{r,j}|\single)$. It is interesting to note that neither Eq.~\eqref{eq:p_binary} nor Eq.~\eqref{eq:p_single} requires the direct comparison between two different observations, as they enter only as independent measurements. This implies that the framework presented here can be straightforwardly applied to single-epoch observations.

Making use of these equation it is possible to draw a parallelism between our method and the one presented in \citet{sana:2013}, where a given object is flagged as a spectroscopic binary if 
\begin{equation}\label{eq:flag_sana}
    \max_{i,j}\qty(\frac{|\hat v_i - \hat v_j|}{\sqrt{\delta_i^2+\delta_j^2}}) > 4\,,
\end{equation}
as per their Eq.~(4): this ensures that the different epochs are likely to be consistent with each other. In addition to this, they also consider also a minimum amplitude threshold,
\begin{equation}
    |\hat v_i - \hat v_j| > C\,,
\end{equation}\label{eq:19}
to account for potential intrinsic variability in the observed radial velocity of the star itself \citep[e.g.][]{2009A&A...507.1585R}.

In this framework, the consistency requirement is encoded in the fact that all the observed posterior distributions $\mathcal{N}(v_{cm}|\hat v_{i,j},\delta_{i,j})$ for a given object, under the assumption of single star, refers to the same quantity $v_{cm}$. Binary components are therefore identified in a probabilistic sense as these objects whose observations are consistent with each other and, at the same time, consistent with the expected thermal radial velocity distribution. This second requirement makes it possible to tell apart single stars and binary components with as little as one single epoch, albeit with limited precision.
The potential intrinsic variability of a single star is not explicitly accounted for in the statistical framework presented in this work but, in presence of multiple epochs and under the assumption of stochastic variability, it is possible to either marginalise over the intrinsic variability or identify such objects as outliers with respect to the underlying CM radial velocity distribution.

In conclusion, making use of the probabilities derived above, it is possible to draw samples from $p(\mathbf{z}|\mathbf{O})$ using the Gibbs sampling scheme and evaluate the probability for each object to be a single star, $\psingle$, and the observed single star fraction, $w_\single$, along with their associated uncertainties.

\subsubsection{Multiple radial velocity populations}
The results presented above assume that the observed stars are all part of the same radial velocity population, i.e. drawn from the same Gaussian distribution, while deriving Eq.~\eqref{eq:p_single}. When applying our method to real observations, it is possible that the observed sample is not homogeneous: it may happen that objects with different radial velocity properties are analysed together.
Whereas for the non-parametric analysis of the radial velocity distribution this issue simply turns into the fact that the reconstructed distribution will show multiple prominent modes, the presence of different sub-populations can significantly affect the identification of binary candidates. This is due to the fact that in our framework the probability of being a single star is linked to the intrinsic radial velocity distribution.

If the non-parametric reconstruction suggests that the intrinsic radial velocity distribution is composed by more than one population, it is possible to include the population membership of each object in the analysis. In the following, we will assume that the number of populations, denoted with $K$, can be decided by looking at the non-parametric reconstruction of the radial velocity distribution: for each of these populations -- labelled with the index $j$ -- the mean velocity $V_{0,j}$ and velocity dispersion $\sigma_j$ are estimated using the same procedure described above.
We introduce here a second set of auxiliary variables, $\mathbf{c} = \{c_1, \ldots, c_N\}$, where $c_i = 1,\ldots,K$ denotes the cluster membership of the $i-$th star. In the same fashion as before, assuming the knowledge of all the values of $\mathbf{c}_{-i}$, the joint probability for $z_i$ and $c_i$ reads
\begin{multline}
    p(z_i = \single/\binary, c_i = j|O_i,\mathbf{z}_{-i}, \mathbf{c}_{-i})\\ \propto p(z_i = \single/\binary|O_i,\mathbf{z}_{-i}, c_i = j)p(c_i = j|O_i,\mathbf{c}_{-i})\,.
\end{multline}
The first term on the right-hand side is the probability described in Eq.~\eqref{eq:singlebinaryprob}, with the only differences that the centre-of-mass radial velocity distribution now depends on the specific parameters of the $j-$th population, $p(v_{cm}|V_{0,j},\sigma_j)$ and $N_\single$ and $N_\binary$ are to be counted considering only these stars for which $c_i = j$.

The additional term $p(c_i = j|O_i,\mathbf{c}_{-i})$ can be rewritten as
\begin{equation}
    p(c_i = j|O_i,\mathbf{c}_{-i}) \propto p(O_i|c_i = j)p(c_i = j|\mathbf{c}_{-i})\,.
\end{equation}
The first term reads
\begin{multline}
    p(O_i|c_i = j) = \int p(O_i|v_{cm})p(v_{cm}|c_i = j) \dd v_{cm}\\ \propto  \int p(v_{cm}|O_i)p(v_{cm}|c_i = j) \dd v_{cm} \,,
\end{multline}
where we assumed a uniform prior on $p(v_{cm})$. The probability $p(v_{cm}|O_i)$ can be approximated with the DPGMM reconstruction used for the non-parametric radial velocity distribution inference, providing a simple way of evaluating this integral.
The second term, $p(c_i = j|\mathbf{c}_{-i})$, denotes the probability of assigning the object to the $j-$th population given the number of already assigned members per population. This probability is derived in the same way as Eq.~\eqref{eq:p_classes_single_binary} simply replacing the symmetric Beta distribution in Eq.~\eqref{eq:betadist} with the symmetric Dirichlet distribution, its multivariate generalisation:
\begin{equation}
    p(\mathbf{q}|\alpha)= \frac{\Gamma(\alpha)}{\qty(\Gamma(\alpha/K))^K}\prod_j q_j^{\alpha/K-1}\,.
\end{equation}
Here $\mathbf{q} = \{q_1,\ldots,q_K\}$ are the relative probabilities of the different populations and $\alpha$ denotes the concentration parameter of the Dirichlet distribution. Taking the marginalisation over $\mathbf{q}$ and denoting with $N_j$ the number of objects associated with the population $j$, we get
\begin{equation}
    p(c_i = j|\mathbf{c}_{-i}) = \frac{N_j + \alpha/K}{N+\alpha}\,.
\end{equation}
This extension to the framework can be useful not only in presence of multiple populations, but also to capture the potential presence of outliers in the observed sample that could contaminate the inference. Modelling the intrinsic radial velocity distribution of a fictitious `outlier' population as a uniform distribution $\mathcal{U}(v_{cm})$, all the objects that do not appear consistent with any of the inferred population will be assigned to this population.

For single-epoch observations, multiple radial velocity populations and candidate binaries have effects that are indistinguishable in the centre-of-mass radial velocity distributions. As shown in the next section, this degeneracy can be broken with multiple observations. In the case of single-epoch observations, we recommend carefully examining the recovered radial velocity distribution to assess the eventual presence of sub-population and include in the analysis only the ones that appear to be robust and populated by a significant number of objects with respect to the total number of analysed stars.

\section{Simulated populations}\label{sec:simulations}
To demonstrate the effectiveness of the methodology presented in this paper, we applied it to two sets of simulated observations: one with a single radial velocity population and another displaying a bimodal radial velocity distribution. They are drawn as following:
\begin{itemize}
    \item The intrinsic radial velocity distribution $p(v_{cm})$ is, given by either a single Gaussian distribution or by a superposition of two Gaussian distributions, depending on the case. Each distribution is characterised in the relevant subsection;
    \item The distribution $p(v_r|v_{cm})$ for stars in binary systems is assumed to be uniformly distributed between $v_{cm}-\Delta v$ and $v_{cm}+\Delta v$, with $\Delta v = 10\ \kms$. In principle, close binary systems can have $\Delta v$ significantly larger than $10\ \kms$: the scenario we selected, however, corresponds to the most challenging case in which the radial velocity variations due to binarity are comparable to those caused by stellar variability. In reality, binary systems will have larger radial velocity variations, and therefore will be easier to identify;
    \item For each star, we realised a random number of simulated epochs (between 3 and 6), drawing for each epoch a $\hat{v}$ value from a Gaussian distribution centred on the radial velocity $v_r$. The measurement uncertainty $\delta$ is assumed to be the largest between $10\%$ of $v_r$ and $0.5\ \kms$.
\end{itemize}
The analysis presented in this Section, as well as the ones presented in the subsequent one, are made using \textsc{raven}\footnote{Publicly available at \url{https://github.com/sterinaldi/raven}.}, a framework built upon \textsc{figaro}\footnote{Publicly available at \url{https://github.com/sterinaldi/FIGARO} and via \href{https://pypi.org/project/figaro/}{\texttt{pip}}.} \citep{rinaldi:2024:figaro}, the inference code designed to reconstruct probability densities with (H)DPGMM.

\subsection{Single radial velocity population}\label{sec:sim_one_pop}
The intrinsic distribution, for this first example, is assumed to be Gaussian with $V_0 = 0\ \kms$ and $\sigma_V = 2.5\ \kms$.
Using the prescriptions given above, we simulated 15 single stars, labelled from S1 to S15, and 14 stars in binary systems (B1-B14).

\begin{figure}
    \centering
    \includegraphics[width=\columnwidth]{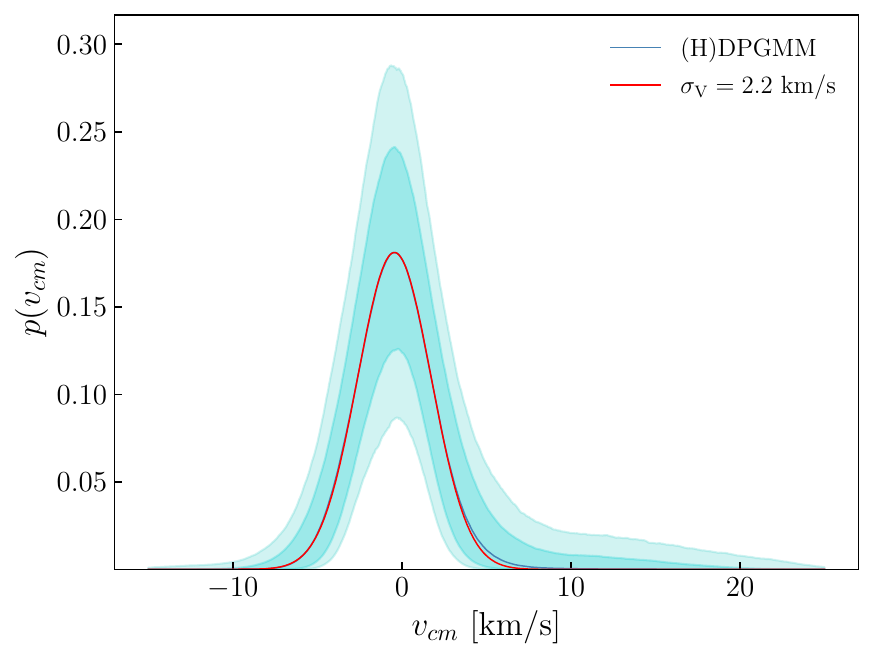}
    \caption{Centre-of-mass radial velocity distribution of the dataset simulated in Sec.~\ref{sec:sim_one_pop} using all the available epochs. The blue line marks the median recovered distribution and the shaded areas correspond to the 68\% and 90\% credible regions. The red line marks the estimated $v_{cm}$ distribution.}
    \label{fig:rv_simulation}
\end{figure}

We applied our methodology to this simulated dataset: the inferred radial velocity distribution is reported in Figure~\ref{fig:rv_simulation}.
The reconstructed distribution is consistent with a Gaussian distribution with $\sigma_V = 2.2\ \kms$, in good agreement with the simulated value of $\sigma_V = 2.5\ \kms$. Due to the specific way in which $\sigma_V$ is obtained -- a point estimate following Eq.~\eqref{eq:jsdistance} -- we cannot associate an uncertainty to this measurement, thus a small disagreement between these quantities is expected.

\begin{figure*}
    \centering
    \subfigure[Multi-epoch]{
    \includegraphics[width=0.98\columnwidth]{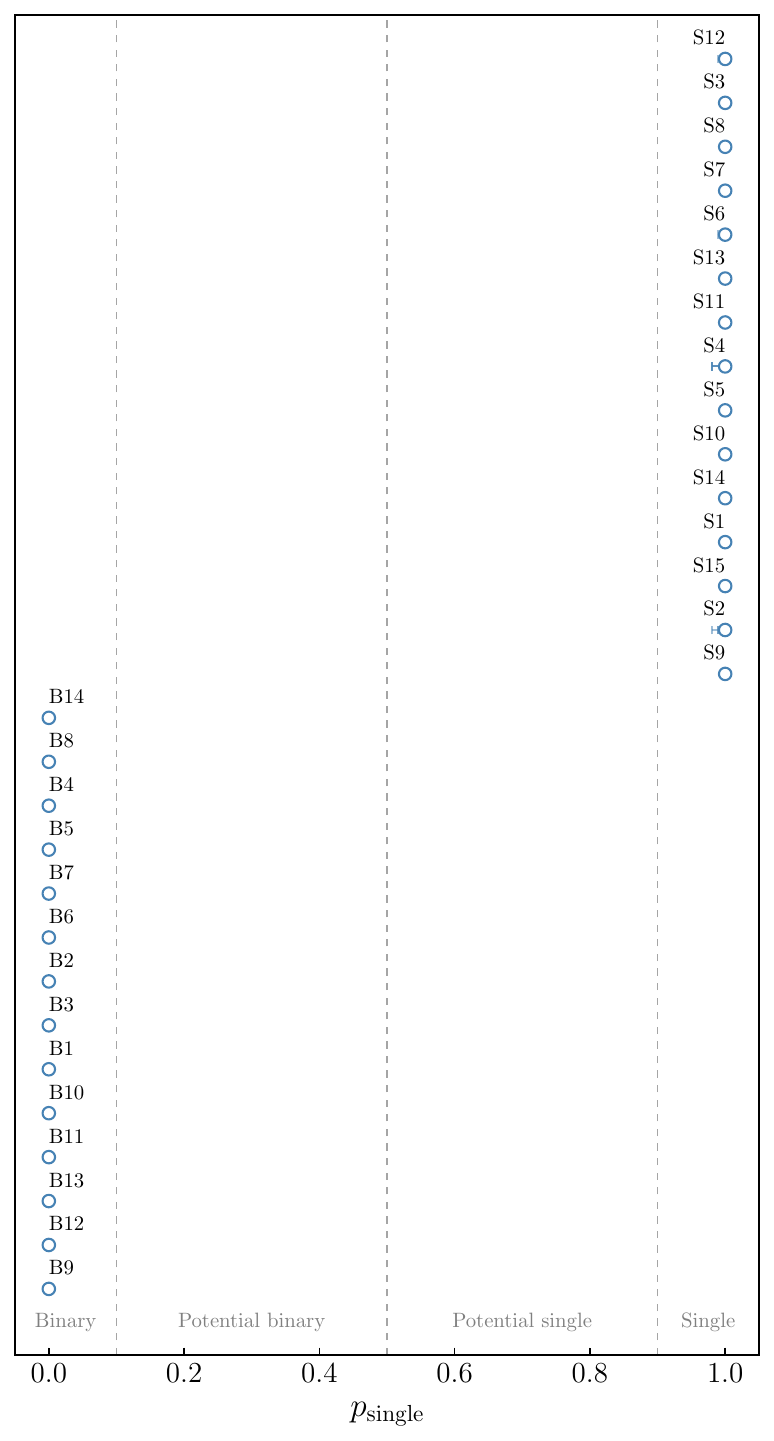}\label{fig:p_single_simulation}
    }
    \subfigure[Single-epoch]{    \includegraphics[width=0.98\columnwidth]{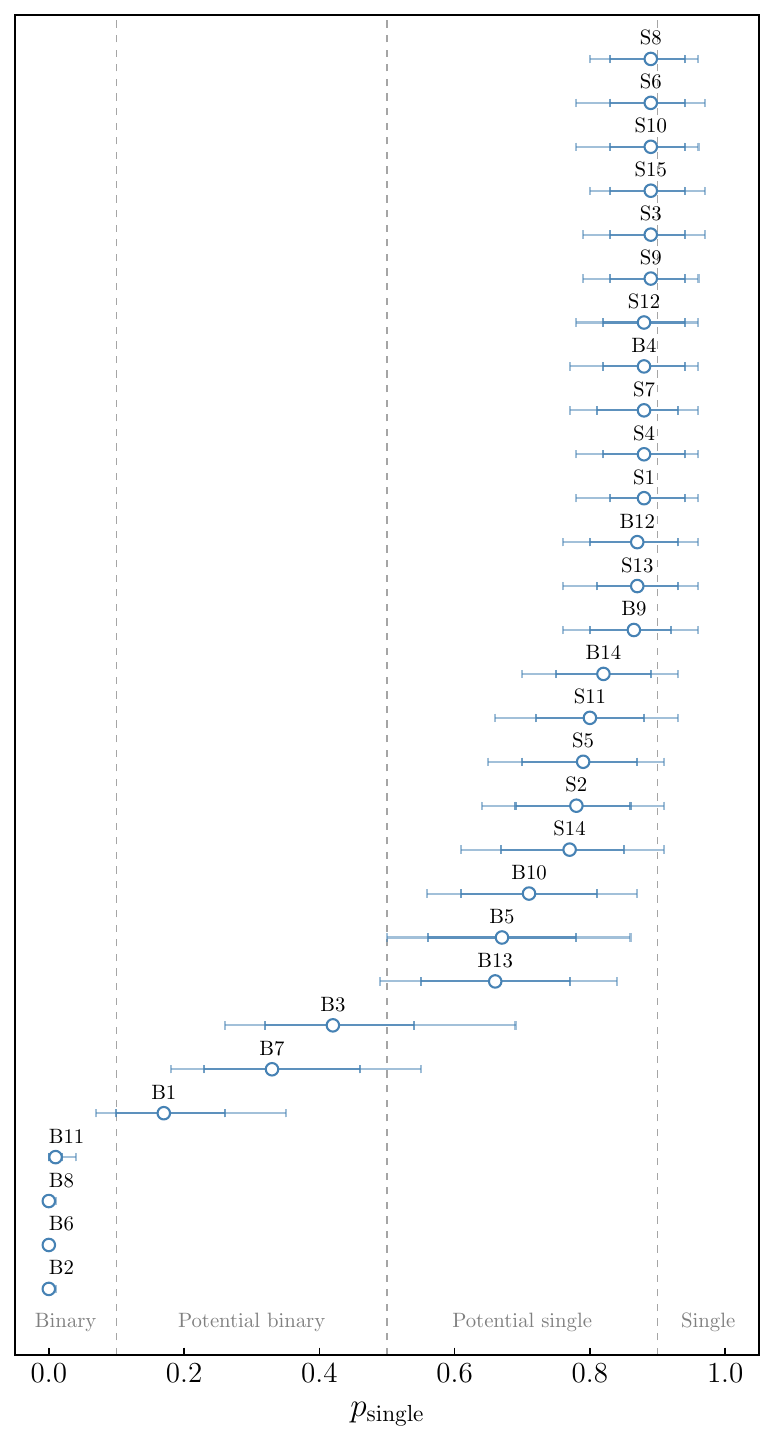}\label{fig:p_single_one_epoch_simulation}
    }
    \caption{Probability of being a single star for each of the 29 objects simulated in Sec.~\ref{sec:sim_one_pop} as evaluated with our framework along with the associated uncertainty (68\% and 90\% credible intervals). Left: multi-epoch analysis including all available measurements. Right: single-epoch analysis including only one randomly selected measurement per star.}
\end{figure*}

We now turn our attention to the problem of categorising the available objects: this can be done using the values of $V_0$ and $\sigma_V$ estimated with the non-parametric analysis. For each object, we evaluated $\psingle$ and its associated uncertainty, reported in Figure~\ref{fig:p_single_simulation}.

Making use of the median value of $\psingle$, we define 4 different classes of objects: 
\begin{itemize}
    \item Confident binary candidate: $\psingle < 0.1$;
    \item Potential binary candidate: $0.1<\psingle<0.5$;
    \item Potential single star: $0.5<\psingle<0.9$;
    \item Confident single star: $\psingle > 0.9$.
\end{itemize}
We opted for the label `binary candidate' to highlight the fact that this framework is not able to distinguish between intrinsic variables and binary components. For real data, where the presence of pulsating stars cannot be excluded a priori, a second and potentially physically informed layer of categorisation, as well as additional epochs if available, should be added a posteriori on a case-by-case basis and applied to these stars that are flagged as binary candidates. Concerning our simulated multi-epoch dataset, all objects are correctly categorised either as confident single stars or confident binary candidates.

An interesting property of the framework we presented in this paper is the fact that it does not specifically requires multi-epoch observations to be applied: the fact that with our methodology we can reconstruct in a non-parametric fashion the expected distribution for single stars allows us to assess the probability for single stars with as little as one epoch per object. 
This is due to the fact that the two probabilities in Eqs.~\eqref{eq:p_binary} and~\eqref{eq:p_single} are factorised as products, thus they can be evaluated with one epoch only. 
Naturally, the more information is available in the form of multiple observations, the more accurate the inference will be: nonetheless, having reliable preliminary results with limited observations would allow for a optimised use of observation time.

To demonstrate this, we built two additional datasets using the 29 simulated objects described above. For each star, we randomly selected one and two epochs respectively: these reduced catalogues are then used to both reconstruct $p(v_r)$ and evaluate $\psingle$ for each object. The different $\psingle$ for the single-epoch catalogue are reported in Figure~\ref{fig:p_single_one_epoch_simulation}. Using our method we are able to correctly flag as either binary or single 23 objects out of 29, $79\%$ of the total, with one epoch only. This figure grows to $93\%$ using two epochs, suggesting that this method can be highly effective even with a limited number of available observations.

\subsection{Bimodal population}\label{sec:sim_two_pop}
In this second example, we consider the case in which the intrinsic radial velocity distribution is composed of two distinct Gaussian populations, labelled $A$ and $B$:
\begin{equation}
    p(v_{cm}) = \frac{1}{2}\mathcal{N}(v_{cm}|V_{0,A}\sigma_{V,A}) + \frac{1}{2}\mathcal{N}(v_{cm}|V_{0,B}\sigma_{V,B})\,.
\end{equation}
Here we set $V_{0,A} = 2.5\ \kms $, $V_{0,B} = -4\ \kms$ and $\sigma_{V,A} = \sigma_{V,B} = 2.5\ \kms$. We opted for a radial velocity separation that is comparable with the velocity dispersion of the two clusters in order to provide a challenging scenario for the framework.
We simulated 20 stars per sub-population (10 binaries and 10 singles) using the recipe outlined at the beginning of this Section. Each object is labelled A-S$N$ for single stars and A-B$N$ for binaries in the population A (B-S$N$ and B-B$N$ for population B). On top of this, we added an additional object (O1) modelled as a single star with $v_{cm} = -21\ \kms$. This object is meant to represent an outlier of the population, such as a background star, that would be flagged as a single star by Eq.~\eqref{eq:flag_sana} but should not be included in the inference of the intrinsic radial velocity distribution of the cluster itself.

\begin{figure*}
    \centering
    \subfigure[All epochs]{
    \includegraphics[width=0.98\columnwidth]{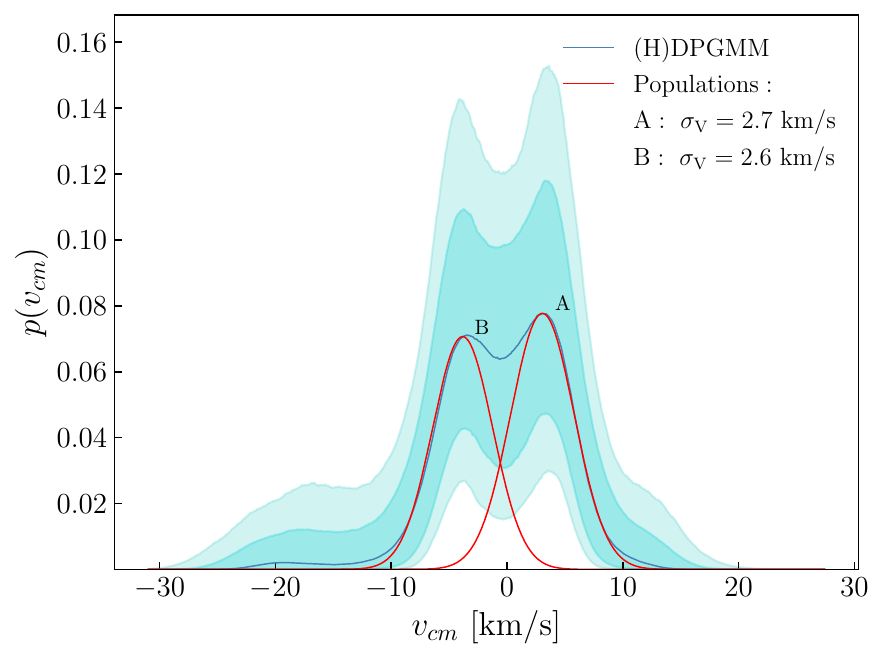}\label{fig:rv_simulation_two_pop}
    }
    \subfigure[One epoch]{
    \includegraphics[width=0.98\columnwidth]{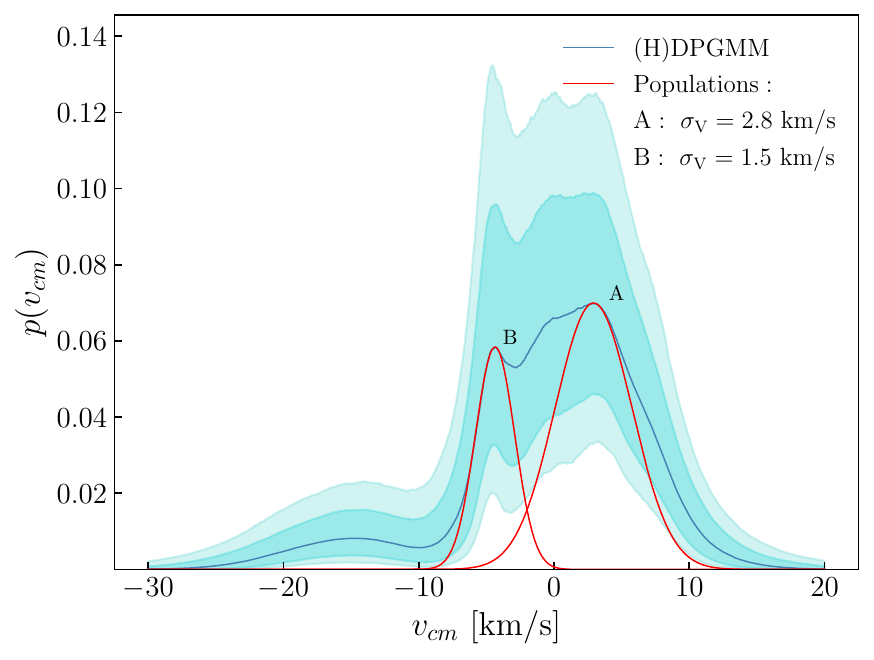}\label{fig:rv_simulation_two_pop_one_epoch}
    }
    \caption{Centre-of-mass radial velocity distribution of the dataset simulated in Sec.~\ref{sec:sim_two_pop}: Left: analysis using all the available epochs. Right: analysis using only one randomly selected epoch. In both cases it is possible to assess the presence of substructures in the radial velocity distribution. The blue line marks the median recovered distribution and the shaded areas correspond to the 68\% and 90\% confidence intervals. The red line marks the two estimated $v_{cm}$ distributions.}
\end{figure*}

The inferred radial velocity distribution obtained using all the available epochs is reported in Figure~\ref{fig:rv_simulation_two_pop}. From the non-parametric reconstruction, we see that the intrinsic distribution has more structure than a simple Gaussian distribution (like the one reported in Fig~\ref{fig:rv_simulation}), hinting towards the presence of two modes in the population.
The two recovered Gaussian sub-populations are consistent with the simulated population. Moreover, the additional feature at around $v_{cm} = -20~\kms$ marks the presence, in the dataset, of an object not consistent with the rest of the population, the outlier O1.

With this second population, we jointly infer the population membership and the probability of being a single star, under the assumption that two populations are present in the dataset plus an additional category for potential outliers. We found that we are able to correctly flag all the objects as either single stars or binaries, including the outlier O1. At the same time, our framework allows to assess the probability for each object to belong to one of the Gaussian populations or to the outlier category. To determine the membership of each object, we considered the median values for $p($A$)$, $p($B$)$ and $p($O$)$ as inferred by our framework, and assigned each object to the population with the largest probability. Doing so, 83\% of the objects were associated with the correct population. For completeness, the inferred membership probabilities for each object are reported in Table~\ref{tab:membership_simulation}.

The same dataset has been also analysed in a single-epoch scenario, randomly selecting one epoch per object.
Figure~\ref{fig:rv_simulation_two_pop_one_epoch} reports the intrinsic distribution we get from the non-parametric analysis: even with as little as one epoch, the recovered distribution hints at the presence of two sub-populations.
In this case, we are able to correctly flag as binary candidates or single stars 26 objects out of 41 (63\%) of the objects.
This reduced performance is due to the fact that, with a single epoch, it is likely that some binary stars belonging to a sub-population might end up having an observed radial velocity consistent with the other sub-population, artificially boosting $\psingle$ for these objects. Therefore, in presence of multiple populations with similar radial velocity distributions (such as the example presented here), care must be taken while interpreting the results.
However, it is important to note that this simulation show a challenging scenario, where the RV amplitude is much lower than what one might expect for massive close binaries. We expect that when applied to real observations, the method performs better even in the single-epoch case.
The degeneracy between binary systems and multiple populations quickly breaks when adding even a single additional epoch: when we included two randomly selected epochs in the analysis, the binary and membership analysis is able to correctly discriminate between singles and binaries for 38 objects out of 41 (93\%) and to properly assign the population membership for 78\% of the simulated stars.

\section{H$_{\mathrm{II}}$ region M17}\label{sec:test_case}

In this section, we apply our methodology to the multi-epoch radial velocity measurements of 20 O and B-type stars in the massive giant H$_\textsc{ii}$ region M17. 18 of these stars have been characterised by \citet{2024A&A...690A.113B}, who determined their physical properties through quantitative spectroscopy. A multi-epoch spectroscopic analysis has recently been done by \citet{ramirez:2024}, where the authors identified the binary stars through multi-epoch spectroscopy using an approach similar to \citet{sana:2013}.
In this Section we will use the same approach to benchmark the results obtained with our inference scheme.
The results presented in this Section can be reproduced using the code stored in the \textsc{raven} repository.

\begin{figure}
    \centering
    \includegraphics[width=\columnwidth]{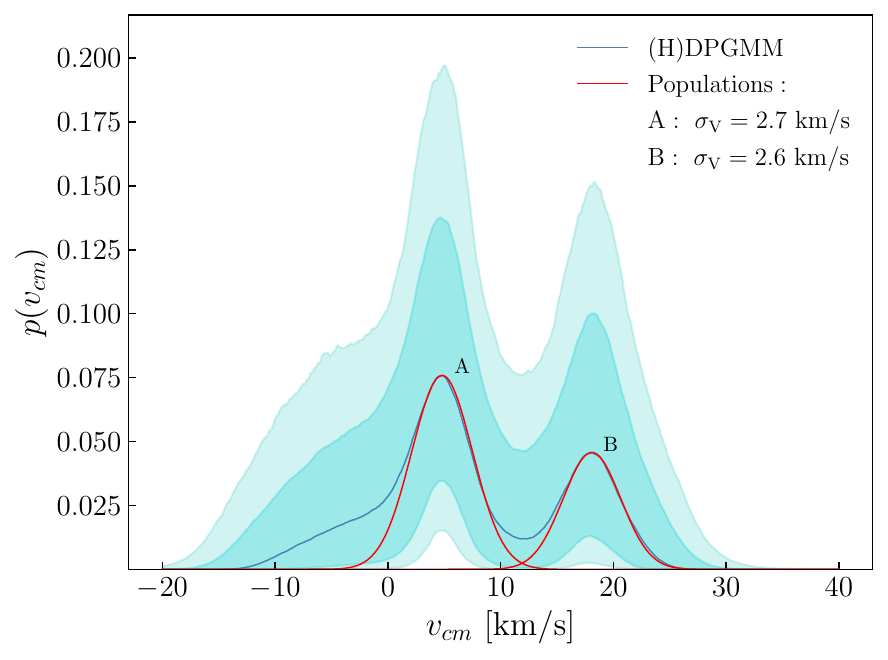}
    \caption{Centre-of-mass radial velocity distribution of the 20 stars from \citet{ramirez:2024} using all the available epochs}. The blue line marks the median recovered distribution and the shaded areas correspond to the 68\% and 90\% credible regions. The red lines marks the two estimated $v_{cm}$ distributions.
    \label{fig:rv_M17}
\end{figure}

Our data set contains radial velocity measurements for 20 stars with a number of epochs ranging from 1 to 5 as indicated in Table~1 of \citet{ramirez:2024}.
Figure~\ref{fig:rv_M17} reports the inferred $v_{cm}$ distribution using all the available measurements. Our reconstruction shows two prominent Gaussian-like features peaking at $V_0 = 4.5\ \kms$ with standard deviation  $\sigma_\mathrm{V} = 2.7\ \kms$ ($A$) and $V_0 = 18.6\ \kms$ with standard deviation  $\sigma_\mathrm{V} = 2.6\ \kms$ ($B$). The additional feature on the side of the Gaussian distributions are due to the presence of binary systems and/or outliers, as described in Section~\ref{sec:methods}.
The radial velocity difference between the two distributions is roughly consistent with the $8.9\ \kms$ velocity dispersion quoted by \citet{ramirez:2024}.

\begin{figure*}
    \centering
    \subfigure[All epochs]{
    \includegraphics[width=0.64\columnwidth]{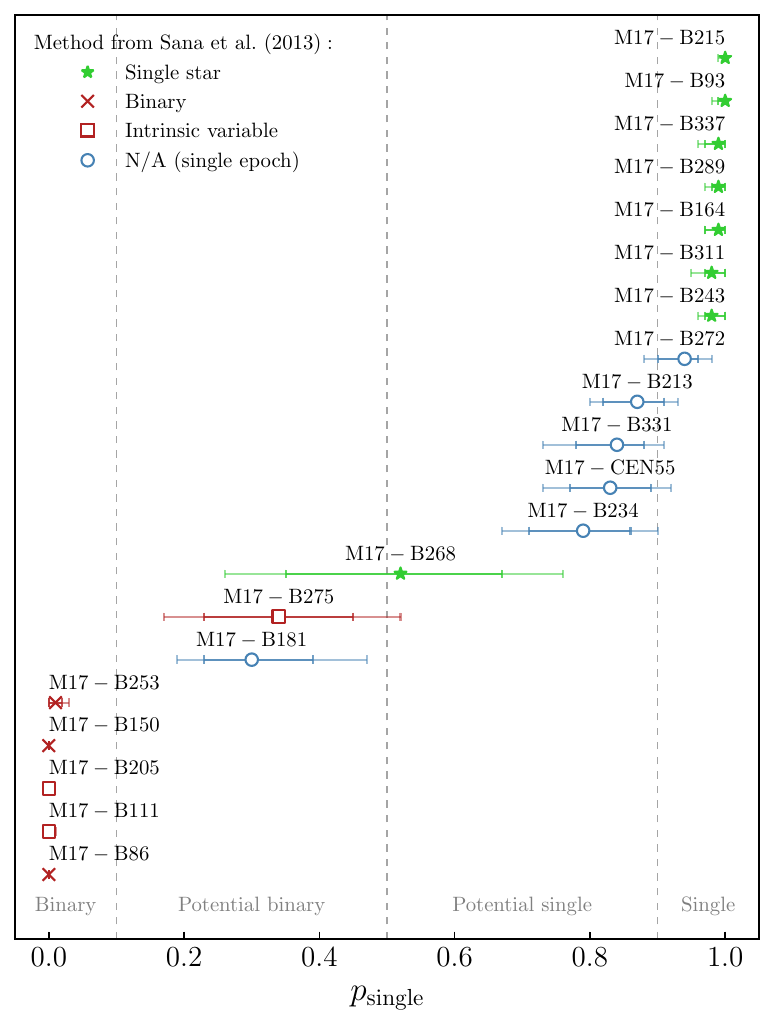}
    \label{fig:p_single_M17}
    }
    \subfigure[One epoch]{
    \includegraphics[width=0.64\columnwidth]{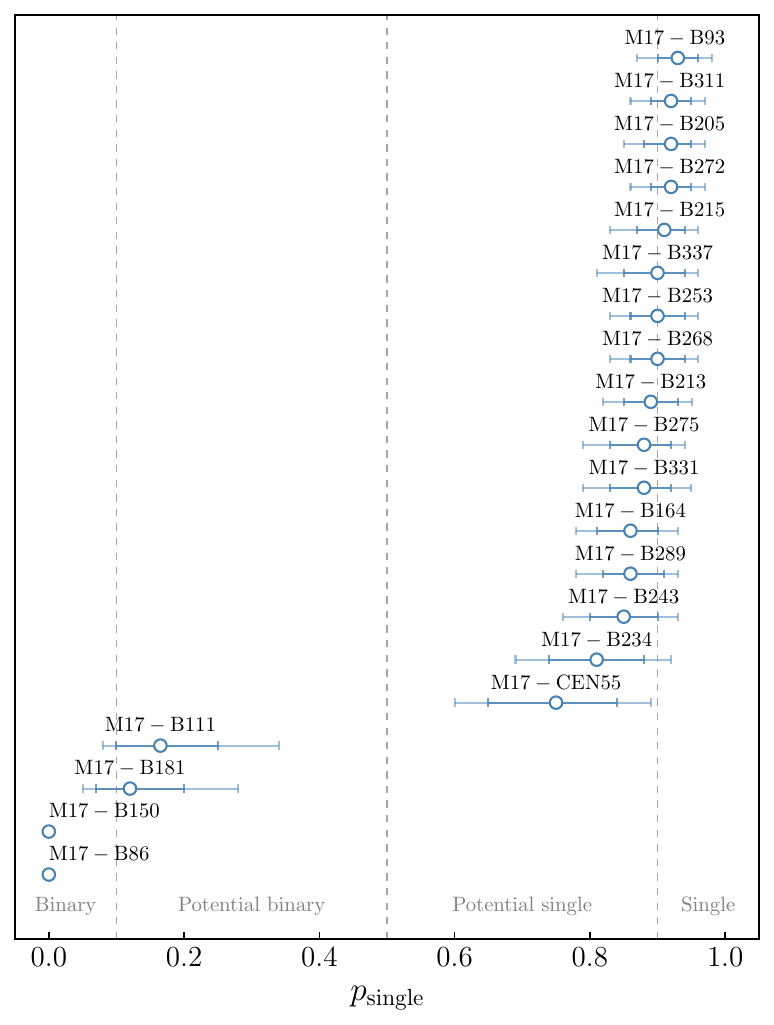}
    \label{fig:p_single_one_epoch_M17}
    }
    \subfigure[Two epochs]{
    \includegraphics[width=0.64\columnwidth]{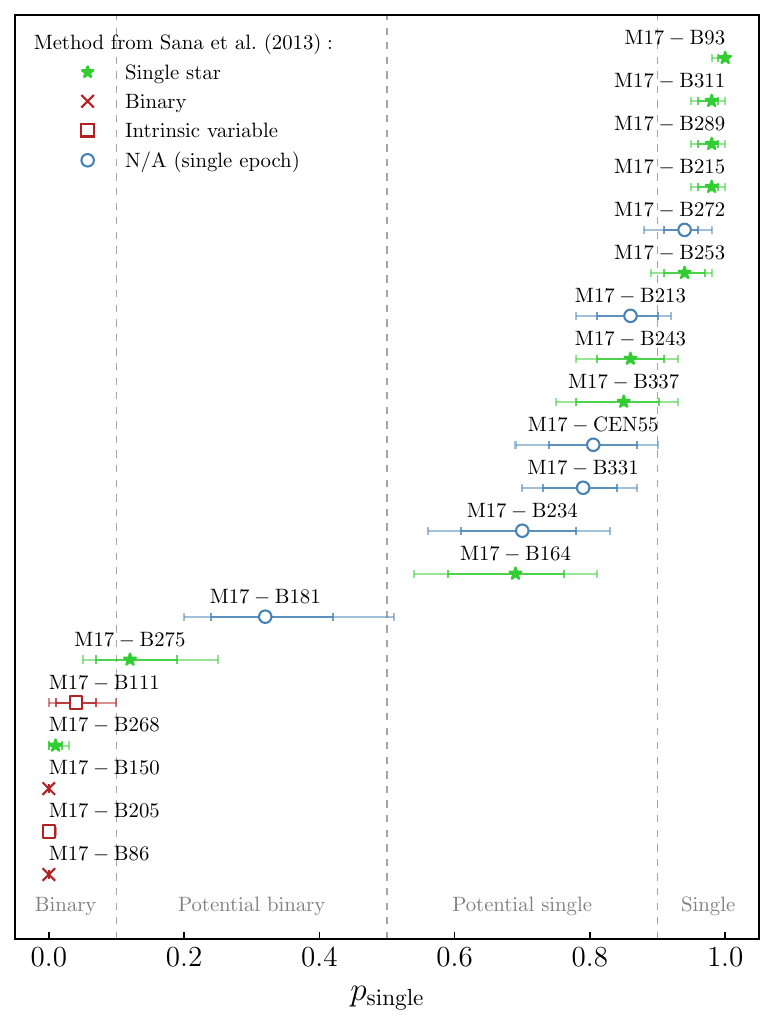}
    \label{fig:p_single_two_epochs_M17}
    }
    \caption{Probability of being a single star for each of the 20 objects considered in this work as evaluated with our framework along with the associated uncertainty (68\% and 90\% credible intervals). From left to right: multi-epoch analysis including all available measurements. Each object is marked according to the label assigned using the method described in \citet{sana:2013}; single-epoch analysis including only one randomly selected measurement per star; analysis including only two randomly selected epochs (when available).}
\end{figure*}

Having characterised the CM velocity distribution, we can focus on the problem of discriminating between binary system components and single stars. Making use of the framework presented in Sec.~\ref{sec:methods} and assuming the presence of two populations as estimated from the non-parametric analysis, we evaluated for each object the probability of being a single star as in Eq.~\eqref{eq:singlebinaryprob}: our results are reported in Figure~\ref{fig:p_single_M17}.

Out of 20 objects, 5 are identified as confident binary candidates and 8 as confident single stars, including B272 which only has single-epoch observations.
Comparing our assignments with the ones obtained with the method presented in \citet{sana:2013}, we find that the two methods are in agreement for all the objects for which multi-epoch observations are available. In particular, the three spectroscopic binaries B86, B150 and B256 are all flagged as confident binaries. Additionally, B181 is flagged as a potential binary despite having only one epoch available.

\subsection{Identification of binaries with single-epoch observations}
To show the reliability of our approach even with single-epoch observations applied to real data, we randomly selected one epoch per source and evaluated $\psingle$ for each star. We report our findings in Figure~\ref{fig:p_single_one_epoch_M17}. Out of 20 objects, 17 share the same label (binary candidate or single star, either confident or potential) as the analysis made with all epochs, which we will use as benchmark here.
We correctly identify two (B86 and B150) our of the three binaries as reported in \citet{ramirez:2024}. The case of the mislabelled binary B253 is discussed below.
From Figure~\ref{fig:p_single_one_epoch_M17} we see that binary candidates are more likely to be confidently identified even with one epoch only due to the fact that they are not consistent with the expected $p(v_{cm})$ for any of the two populations.
On the other hand, since the uncertainty associated with individual epochs is often comparable with the width of the expected CM velocity distribution, stars with $\psingle > 0.9$ have larger uncertainties with respect to the multi-epoch analysis. 
The fact that the reconstructed radial velocity distribution shows more than one prominent feature (the two populations marked $A$ and $B$ in Figure~\ref{fig:rv_M17} are present also in the single-epoch analysis) can affect the recovered $\psingle$, given that both binary stars and the presence of multiple statistical populations induce the same effect on the observed radial velocities.

This degeneracy is broken making use of multi-epoch observations: to show this, we repeated the exercise of using only a subset of the available epochs, this time randomly selecting two epochs (when available). The inferred $\psingle$ for the 20 objects are reported in Figure~\ref{fig:p_single_two_epochs_M17}. We found that almost all the objects for which multi-epoch observations are available are flagged in agreement to \citet{ramirez:2024}. The exceptions are:
\begin{itemize}
    \item B275: when including all the available epochs this object is an intrinsic variable star. With the two selected epochs, the method presented in \citet{sana:2013} would have classified it as a single, non-variable star. Based on the same two epochs, our method is capable of identifying it as an outlier or `binary candidate';
    \item B268: in the multi-epoch analysis, this star is identified as a single star by both methods (albeit the classification is very uncertain with our framework). With two epochs, our method classifies this object as a binary: this is due to the fact that, for this object, the majority of the available epochs are in the tail of the Gaussian distribution used to model the intrinsic radial velocity distribution;
    \item B253: this object, known to be a binary star, in the single-epoch analysis is flagged as a single star. This is due to the fact that the epoch we selected for the analysis ($4\pm 4\ \kms$) is fully consistent with the main feature of the $v_{cm}$ distribution -- population $A$.
    This mislabelling is consistent with the caveat of our method concerning the single-epoch analysis.
\end{itemize}
Comparing the single/binary classification we get using two epochs only with the analysis done with all the available epochs, we find that the labels are in agreement 95\% of the times.

It is important, once again, to keep in mind that in the case of single-epoch analysis, assuming multiple populations may yield degeneracies with binaries or variable stars, thus care must be taken while drawing strong conclusions based on such data.

\section{Summary}\label{sec:conclusion}
In this paper we present a framework based on Bayesian non-parametric methods to effectively identify spectroscopic binary candidates with as little as one epoch observations of a given cluster. Our method has the potential to identify the sources that are worth following up, before doing multi-epoch observations of the whole population in order to optimise the requested observing time, especially for spectrographs that observe a single object at the time: in a nutshell, the idea behind the method we propose in this paper is to look for observations deviating from an expected intrinsic distribution -- in our case, the Gaussian distribution induced by the cluster dynamics -- and, in case of multi-epoch observations, inconsistencies among subsequent measurements.
This method can be a viable alternative and complementary to other existing methods, such as the one presented in \citet{sana:2013}, which relies on the consistency among epochs only and here used as benchmark, or the ones intrinsically reliant on specific prescriptions for both the intrinsic radial velocity distribution and the velocity distribution in binaries $p(v_r|v_{cm})$, such as the one presented in \citet{cottaar:2014}. With respect to the latter, our approach has two advantages: being able to easily extract information also from the stars that are classified as binary candidates via the hierarchical scheme and not relying on any specific functional form for $p(v_r|v_{cm})$ and $p(v_{cm})$. This flexibility will allow our model, if applied to sufficiently precise measurements, to reveal the presence of substructures in the intrinsic radial velocity distribution.

We test our approach by comparing our results to those presented by \citet{ramirez:2024}, who identified binaries in the M17 region through multi-epoch spectroscopy. We find a perfect match with the objects reported in \citet{ramirez:2024} when our method is applied to multi-epoch observations. The method presented in this paper also has the flexibility be applied to single-epoch observations, finding one potential new binary candidate (B181). 

In this framework, the potential mislabelling of a single star can be due to the object being either intrinsically variable (thus making it statistically indistinguishable from a binary candidate) or having a median radial velocity measurement inconsistent with the expected Gaussian distribution(s). In both cases, our method is able to point them as `noteworthy', leaving us with the task of further investigating the objects' nature.

Overall, we believe that the method we present in this paper, being able to identify potential binary candidates with as little as one epoch only and to flexibly reconstruct the intrinsic radial velocity distribution with minimal assumptions, will be of great help both to optimise the usage of the available observing time and to deepen our understanding of stellar multiplicity in clusters.

\begin{acknowledgements}
The authors are grateful to Marco~Monaci and the anonymous referee for useful discussions and suggestions.
SR acknowledges financial support from the European Research Council for the ERC Consolidator grant DEMOBLACK, under contract no. 770017, and from the German Excellence Strategy via the Heidelberg Cluster of Excellence (EXC 2181 - 390900948) STRUCTURES. 
MCRT acknowledges support by the German Aerospace Center (DLR) and the Federal Ministry for Economic Affairs and Energy (BMWi) through program 50OR2314 ‘Physics and Chemistry of Planet-forming disks in extreme environments’. 
This research made use of the bwForCluster Helix: the authors acknowledge support by the state of Baden-Württemberg through bwHPC and the German Research Foundation (DFG) through grant INST 35/1597-1 FUGG.
\end{acknowledgements}
\bibliographystyle{aa}
\bibliography{bibliography.bib}
\appendix
\onecolumn
\section{Membership probability for the bimodal population simulations}
\input{membership_table.tex}

\end{document}

%% file: membership_table.tex
\begin{table*}[h!]
    \caption{Inferred membership probability for each object included in the simulated catalogue of Section~\ref{sec:sim_two_pop}.}
    \centering
    \begin{tabular}{c|ccc|ccc}
    & & All epochs & & & Two epochs & \\
    \toprule
    Name & $p($O$)$ & $p($A$)$ & $p($B$)$ & $p($O$)$ & $p($A$)$ & $p($B$)$ \\
    \midrule
A-S1&$0.08^{+0.03}_{-0.03}$&$0.89^{+0.03}_{-0.04}$&$0.03^{+0.02}_{-0.01}$&$0.12^{+0.03}_{-0.03}$&$0.82^{+0.04}_{-0.04}$&$0.06^{+0.03}_{-0.02}$\\
A-S2&$0.06^{+0.03}_{-0.02}$&$0.72^{+0.04}_{-0.05}$&$0.22^{+0.04}_{-0.05}$&$0.09^{+0.03}_{-0.03}$&$0.61^{+0.05}_{-0.04}$&$0.30^{+0.04}_{-0.05}$\\
A-S3&$0.13^{+0.04}_{-0.03}$&$0.86^{+0.03}_{-0.04}$&$0.00^{+0.01}_{-0.00}$&$0.19^{+0.04}_{-0.04}$&$0.79^{+0.04}_{-0.04}$&$0.02^{+0.01}_{-0.02}$\\
A-S4&$0.07^{+0.02}_{-0.03}$&$0.76^{+0.04}_{-0.04}$&$0.18^{+0.03}_{-0.04}$&$0.09^{+0.03}_{-0.03}$&$0.70^{+0.05}_{-0.05}$&$0.21^{+0.04}_{-0.04}$\\
A-S5&$0.06^{+0.03}_{-0.02}$&$0.75^{+0.04}_{-0.05}$&$0.18^{+0.04}_{-0.03}$&$0.09^{+0.03}_{-0.03}$&$0.68^{+0.04}_{-0.05}$&$0.23^{+0.04}_{-0.04}$\\
A-S6&$0.06^{+0.03}_{-0.02}$&$0.21^{+0.04}_{-0.04}$&$0.72^{+0.05}_{-0.04}$&$0.09^{+0.03}_{-0.03}$&$0.25^{+0.04}_{-0.04}$&$0.66^{+0.05}_{-0.05}$\\
A-S7&$0.07^{+0.03}_{-0.02}$&$0.88^{+0.03}_{-0.03}$&$0.04^{+0.02}_{-0.02}$&$0.11^{+0.03}_{-0.04}$&$0.81^{+0.03}_{-0.04}$&$0.09^{+0.03}_{-0.03}$\\
A-S8&$0.07^{+0.02}_{-0.02}$&$0.82^{+0.03}_{-0.04}$&$0.11^{+0.04}_{-0.03}$&$0.10^{+0.03}_{-0.03}$&$0.72^{+0.04}_{-0.05}$&$0.19^{+0.03}_{-0.04}$\\
A-S9&$0.09^{+0.03}_{-0.03}$&$0.89^{+0.03}_{-0.03}$&$0.02^{+0.01}_{-0.01}$&$0.13^{+0.04}_{-0.03}$&$0.83^{+0.04}_{-0.03}$&$0.03^{+0.02}_{-0.02}$\\
A-S10&$0.08^{+0.03}_{-0.03}$&$0.89^{+0.03}_{-0.03}$&$0.03^{+0.02}_{-0.01}$&$0.12^{+0.03}_{-0.03}$&$0.83^{+0.03}_{-0.04}$&$0.06^{+0.02}_{-0.03}$\\
A-B1&$0.10^{+0.03}_{-0.03}$&$0.48^{+0.05}_{-0.05}$&$0.42^{+0.05}_{-0.05}$&$0.18^{+0.04}_{-0.03}$&$0.71^{+0.04}_{-0.05}$&$0.11^{+0.03}_{-0.03}$\\
A-B2&$0.20^{+0.05}_{-0.03}$&$0.77^{+0.04}_{-0.04}$&$0.02^{+0.02}_{-0.01}$&$0.21^{+0.04}_{-0.04}$&$0.74^{+0.04}_{-0.05}$&$0.05^{+0.03}_{-0.02}$\\
A-B3&$0.10^{+0.03}_{-0.03}$&$0.41^{+0.04}_{-0.05}$&$0.49^{+0.05}_{-0.05}$&$0.14^{+0.04}_{-0.03}$&$0.33^{+0.05}_{-0.04}$&$0.52^{+0.05}_{-0.05}$\\
A-B4&$0.07^{+0.02}_{-0.03}$&$0.26^{+0.05}_{-0.04}$&$0.67^{+0.04}_{-0.05}$&$0.10^{+0.03}_{-0.03}$&$0.17^{+0.04}_{-0.04}$&$0.73^{+0.05}_{-0.04}$\\
A-B5&$0.17^{+0.04}_{-0.03}$&$0.69^{+0.05}_{-0.04}$&$0.13^{+0.03}_{-0.03}$&$0.16^{+0.04}_{-0.03}$&$0.68^{+0.04}_{-0.06}$&$0.16^{+0.04}_{-0.04}$\\
A-B6&$0.11^{+0.03}_{-0.03}$&$0.78^{+0.04}_{-0.04}$&$0.11^{+0.03}_{-0.03}$&$0.21^{+0.04}_{-0.04}$&$0.78^{+0.04}_{-0.04}$&$0.01^{+0.01}_{-0.01}$\\
A-B7&$0.13^{+0.04}_{-0.03}$&$0.83^{+0.04}_{-0.03}$&$0.03^{+0.02}_{-0.01}$&$0.12^{+0.03}_{-0.03}$&$0.83^{+0.04}_{-0.03}$&$0.04^{+0.03}_{-0.01}$\\
A-B8&$0.10^{+0.03}_{-0.03}$&$0.72^{+0.05}_{-0.04}$&$0.18^{+0.04}_{-0.04}$&$0.15^{+0.04}_{-0.04}$&$0.56^{+0.05}_{-0.05}$&$0.29^{+0.04}_{-0.04}$\\
A-B9&$0.41^{+0.05}_{-0.04}$&$0.58^{+0.04}_{-0.05}$&$0.01^{+0.01}_{-0.01}$&$0.38^{+0.05}_{-0.05}$&$0.61^{+0.04}_{-0.06}$&$0.01^{+0.02}_{-0.01}$\\
A-B10&$0.08^{+0.03}_{-0.03}$&$0.39^{+0.05}_{-0.04}$&$0.53^{+0.04}_{-0.05}$&$0.13^{+0.04}_{-0.03}$&$0.79^{+0.04}_{-0.04}$&$0.07^{+0.03}_{-0.02}$\\
\midrule
B-S1&$0.07^{+0.03}_{-0.02}$&$0.07^{+0.03}_{-0.02}$&$0.86^{+0.03}_{-0.04}$&$0.11^{+0.04}_{-0.03}$&$0.11^{+0.04}_{-0.03}$&$0.77^{+0.04}_{-0.04}$\\
B-S2&$0.06^{+0.03}_{-0.02}$&$0.40^{+0.05}_{-0.05}$&$0.53^{+0.05}_{-0.05}$&$0.09^{+0.02}_{-0.03}$&$0.33^{+0.05}_{-0.04}$&$0.58^{+0.05}_{-0.05}$\\
B-S3&$0.07^{+0.03}_{-0.02}$&$0.06^{+0.02}_{-0.02}$&$0.87^{+0.03}_{-0.04}$&$0.12^{+0.03}_{-0.03}$&$0.11^{+0.03}_{-0.03}$&$0.77^{+0.04}_{-0.04}$\\
B-S4&$0.06^{+0.03}_{-0.02}$&$0.34^{+0.05}_{-0.04}$&$0.59^{+0.04}_{-0.05}$&$0.08^{+0.04}_{-0.02}$&$0.36^{+0.04}_{-0.05}$&$0.55^{+0.05}_{-0.04}$\\
B-S5&$0.08^{+0.03}_{-0.03}$&$0.05^{+0.02}_{-0.02}$&$0.87^{+0.04}_{-0.03}$&$0.13^{+0.03}_{-0.04}$&$0.09^{+0.03}_{-0.03}$&$0.78^{+0.04}_{-0.04}$\\
B-S6&$0.06^{+0.03}_{-0.02}$&$0.47^{+0.04}_{-0.06}$&$0.47^{+0.05}_{-0.05}$&$0.08^{+0.03}_{-0.02}$&$0.44^{+0.05}_{-0.05}$&$0.47^{+0.06}_{-0.04}$\\
B-S7&$0.07^{+0.02}_{-0.03}$&$0.18^{+0.04}_{-0.04}$&$0.75^{+0.04}_{-0.04}$&$0.09^{+0.03}_{-0.02}$&$0.20^{+0.04}_{-0.04}$&$0.70^{+0.05}_{-0.04}$\\
B-S8&$0.12^{+0.04}_{-0.03}$&$0.01^{+0.02}_{-0.01}$&$0.86^{+0.04}_{-0.03}$&$0.19^{+0.04}_{-0.04}$&$0.05^{+0.03}_{-0.02}$&$0.75^{+0.04}_{-0.04}$\\
B-S9&$0.07^{+0.03}_{-0.02}$&$0.09^{+0.03}_{-0.03}$&$0.84^{+0.03}_{-0.04}$&$0.11^{+0.03}_{-0.03}$&$0.13^{+0.04}_{-0.03}$&$0.76^{+0.04}_{-0.04}$\\
B-S10&$0.06^{+0.03}_{-0.02}$&$0.58^{+0.05}_{-0.05}$&$0.35^{+0.05}_{-0.04}$&$0.09^{+0.03}_{-0.03}$&$0.58^{+0.05}_{-0.05}$&$0.34^{+0.05}_{-0.05}$\\
B-B1&$0.10^{+0.03}_{-0.02}$&$0.42^{+0.05}_{-0.04}$&$0.47^{+0.05}_{-0.05}$&$0.11^{+0.04}_{-0.03}$&$0.44^{+0.05}_{-0.05}$&$0.45^{+0.05}_{-0.05}$\\
B-B2&$0.08^{+0.03}_{-0.03}$&$0.72^{+0.04}_{-0.04}$&$0.20^{+0.04}_{-0.04}$&$0.11^{+0.03}_{-0.03}$&$0.62^{+0.05}_{-0.05}$&$0.27^{+0.04}_{-0.04}$\\
B-B3&$0.10^{+0.04}_{-0.03}$&$0.51^{+0.04}_{-0.05}$&$0.39^{+0.05}_{-0.05}$&$0.12^{+0.04}_{-0.03}$&$0.52^{+0.05}_{-0.05}$&$0.35^{+0.05}_{-0.04}$\\
B-B4&$0.20^{+0.05}_{-0.03}$&$0.08^{+0.03}_{-0.03}$&$0.72^{+0.04}_{-0.05}$&$0.21^{+0.04}_{-0.04}$&$0.10^{+0.03}_{-0.03}$&$0.69^{+0.04}_{-0.05}$\\
B-B5&$0.20^{+0.05}_{-0.04}$&$0.03^{+0.02}_{-0.01}$&$0.76^{+0.04}_{-0.04}$&$0.39^{+0.06}_{-0.04}$&$0.05^{+0.02}_{-0.02}$&$0.55^{+0.06}_{-0.05}$\\
B-B6&$0.08^{+0.03}_{-0.03}$&$0.18^{+0.04}_{-0.04}$&$0.74^{+0.04}_{-0.05}$&$0.11^{+0.03}_{-0.03}$&$0.28^{+0.04}_{-0.05}$&$0.62^{+0.05}_{-0.05}$\\
B-B7&$0.47^{+0.06}_{-0.05}$&$0.01^{+0.01}_{-0.01}$&$0.52^{+0.05}_{-0.06}$&$0.52^{+0.05}_{-0.05}$&$0.03^{+0.02}_{-0.01}$&$0.45^{+0.05}_{-0.05}$\\
B-B8&$0.24^{+0.04}_{-0.04}$&$0.01^{+0.02}_{-0.01}$&$0.74^{+0.05}_{-0.04}$&$0.87^{+0.04}_{-0.03}$&$0.01^{+0.01}_{-0.01}$&$0.11^{+0.04}_{-0.03}$\\
B-B9&$0.12^{+0.03}_{-0.03}$&$0.42^{+0.06}_{-0.04}$&$0.46^{+0.04}_{-0.05}$&$0.13^{+0.03}_{-0.03}$&$0.52^{+0.05}_{-0.05}$&$0.35^{+0.05}_{-0.05}$\\
B-B10&$0.13^{+0.05}_{-0.05}$&$0.04^{+0.02}_{-0.02}$&$0.83^{+0.05}_{-0.05}$&$0.46^{+0.05}_{-0.05}$&$0.04^{+0.02}_{-0.02}$&$0.50^{+0.05}_{-0.05}$\\
\midrule
O1&$0.96^{+0.02}_{-0.02}$&$0.00^{+0.01}_{-0.00}$&$0.04^{+0.02}_{-0.02}$&$0.99^{+0.01}_{-0.01}$&$0.00^{+0.00}_{-0.00}$&$0.01^{+0.01}_{-0.01}$\\
    \bottomrule
    \end{tabular}
    \label{tab:membership_simulation}
\end{table*}